\documentclass[
preprintnumbers,
amsmath,amssymb,aps]{revtex4}
\usepackage[colorlinks=true, pdfstartview=FitV, linkcolor=green, citecolor=blue, urlcolor=red]{hyperref}
\usepackage{amssymb}
\usepackage{latexsym}
\usepackage{epsfig}
\usepackage{caption}
\usepackage{subcaption}
\usepackage[normalem]{ulem} 
\def\bre{Barrow entropy}
\def\brp{fractal correction parameter}
\usepackage{color}
\allowdisplaybreaks
\begin{document}
\title{Thermodynamic analysis of
black holes with cloud of strings and quintessence via Barrow
entropy}
\author{Usman Zafar $^{1}$ \footnote{zafarusman494@gmail.com,
usmanzafar366@yahoo.com}, Kazuharu Bamba $^{1}$
\footnote{bamba@sss.fukushima-u.ac.jp}, Tabinda Rasheed $^{2}$
\footnote{tabindarasheed00@gmail.com},  and Krishnakanta
Bhattacharya$^{3}$ \footnote{krishnakanta@dubai.bits-pilani.ac.in}}
\address{$^1$ Faculty of Symbiotic Systems Science, Fukushima University,
Fukushima 960-1296, Japan.\\
$^2$ School of Mathematics and Statistics, Nanjing University of Information Science and Technology, China.\\
$^3$ Department of General Science, Birla Institute of Technology and Science, Pilani, Dubai Campus, Dubai International Academic City, Dubai, United Arab Emirates}
\date{}

\begin{abstract}

We explore a Reissner-Nordstr\"{o}m Anti-de Sitter (RN-AdS) black hole with a cloud of
string and quintessence to study thermodynamics 
and thermodynamic topology
in the presence of Barrow entropy, which is currently being used widely as the horizon of a black hole may not be a smooth surface as described in classical general relativity but instead could have a more intricate fractal-like structure. Here, we study the impact of the fractal correction parameter of Barrow entropy on the thermodynamics
of such BHs.
We compute the first law of black hole thermodynamics and the Smarr relation for RN-AdS black hole with a cloud of
string and quintessence in terms of \bre~ by employing the generalized formula for spherically symmetric spacetime which is directly derived from the Einstein field equation. The significance of Barrow entropy has been verified from thermodynamic topology as well. We also found that the non-zero topological charge indicates the presence of the critical point. 

\end{abstract}

\maketitle

\section{Introduction}
Black holes (BHs) are among the most fascinating objects in the universe. The existence of these objects can be predicted from Einstein's GR, which is currently the best-known theory for humankind for understanding gravitational force. Over the years, studying different BHs has opened new windows to obtain a new perspective on gravitational force. As a result, people started studying various BHs with multiple features. One such example is the study of Letelier, which focuses on solutions to the Einstein
equations for a BH surrounded by the strings cloud and quintessence (SSCQ)
\cite{Letelier:1979ej}. This approach considers nature to be composed of extended
objects like strings instead of point particles. Letelier's work
generalizes the Schwarzschild solution by incorporating a
spherically symmetric cloud of strings, modifying the gravitational
field. If the BH's mass is zero, only a cloud of strings
remains, resulting in a naked singularity at the origin \cite{Letelier:1979ej}.

BH thermodynamics has become a vital field in gravitational
physics since Bekenstein and Hawking's 1970s work, shedding light on
quantum mechanics and gravity. Researchers have linked BHs to liquid-gas
systems by studying the thermodynamics of charged AdS BHs
\cite{Chamblin:1999tk, Chamblin:1999hg}, revealing phase transitions analogous to Van
der Waals fluids (for more details check these
\cite{Sekiwa:2006qj,Caldarelli:1999xj,Kastor:2009wy,Kubiznak:2012wp,Cai:2013qga, Banerjee:2012zm}).
$P-V$ criticality and
phase transition for various complex BHs has been addressed in Refs.~\cite{Li:2018rpk,Mo:2013ela,Xu:2015rfa,Momeni:2016qfv,Karch:2015rpa,Mancilla:2024spp, Fernando:2016sps}). There are numerous methods to discuss phase transition phenomena or critical points like thermodynamic geometry and topology. There are some problems with the thermodynamic geometry because sometimes we have a complex thermodynamic potential, making it quite difficult to study the phase transition or zero point using the thermodynamic geometry formalism. Since in our case, we have additional fields, it is quite difficult to study the zero point or critical point for RN-AdS BH SSCQ. Therefore, we employ thermodynamic topology to study the critical or zero points for the BHs with additional fields. Most recently, thermodynamic topology has been widely used to discuss the phase transition or critical points without analyzing the thermodynamic properties and field equations (for more details, see Refs.~\cite{Wei:2020rbh, Wei:2022dzw,Wei:2021vdx,Yerra:2022alz,Yerra:2022eov,Gogoi:2023xzy,Gogoi:2023qku,Yerra:2022coh,Wei:2022mzv}). Thermodynamics of RN-AdS BH SSCQ has been discussed with the help of the Hawking-Bekenstein entropy (for further details, check these Refs.~\cite{deMToledo:2018tjq,Toledo:2019amt,Bekenstein:1973ur,Hawking:1974rv, Hawking:1976de}). However, analyzing the  BHs having additional fields by using the Hawking-Bekenstein entropy seems to have some limitations \cite{Rani:2022xza, Jawad:2022lww, Capozziello:2025axh,Jawad:2020ihz}. These additional fields complicate the thermodynamic characteristics such as phase transition analysis and stability conditions \cite{deMToledo:2018tjq, Toledo:2019amt}. However, \bre~ introduces the concept of fractal geometry on BH's horizon, which incorporates a gravitational quantum effect \cite{Barrow:2020tzx}. Thus, it provides a detailed mathematical framework that may produce a novel thermodynamic behavior and phase transition of BHs. Moreover, \bre~  offers a more generic approach to understanding the BH thermodynamics in the existence of additional fields.


Investigating this area could provide new insights into the phase transitions and thermodynamic behavior of these unique BHs, filling a significant gap in the current research and advancing our understanding of BH thermodynamics.
The entire paper
is presented in the following step-by-step fashion. A Review of the RN-AdS BH SSCQ  is discussed in
Sec.~\ref{sec-2}. Thermodynamic properties of
RN-AdS BH SSCQ by \bre~ is covered in Sec.~\ref{sec-3}. We have discussed the impact of the intricacy of the horizon structure on the thermodynamics of RN-AdS BH SSCQ. In
addition, we revisit the first BH law of thermodynamics and construct the Smarr relation, which we computed from generalized formulations by
considering extensive parameters. Moreover, the thermodynamic topology of RN-AdS BH SSCQ in terms of \bre~ is discussed for $\Delta=0$, which indicates that horizon structure is smooth, and $\Delta=1$, which indicates that horizon has a most intricate structure in Sec.~\ref{sec-5}. Lastly, Sec.~\ref{sec-6} has detailed conclusions of this work.

\section{Reissner-Nordstr\"{o}m Anti-de Sitter Black Holes Surrounded By String Clouds and Quintessence: A Review}\label{sec-2}

The cloud of strings affects the horizon structure by enlarging the
event horizon radius and introducing a solid deficit angle, similar
to the spacetime of a global monopole \cite{Barriola:1989hx}. This influence
suggests significant astrophysical implications, motivating further
investigation into the role of the cloud of strings. The discovery
of the universe's accelerated expansion is a significant achievement in
observational cosmology \cite{SupernovaCosmologyProject:1998vns,Planck:2018vyg,Planck:2018jri}, indicating the presence
of a gravitational repulsive energy component with negative
pressure. This negative pressure might originate from a uniform,
time-independent fluid known as quintessence or quintessence dark
energy (Many ideas have been suggested to construct
feasible models of dark energy; these are summarized in Refs.~\cite{Cardenas:2002np,Guo:2004fq,Copeland:2006wr,Xia:2006rr} and for better understanding of dark energy, check the reviews on dark energy in Refs.~\cite{Padmanabhan:2007xy,Durrer:2007re, Bamba:2012cp,Joyce:2014kja,Frusciante:2019xia}). Quintessence has an energy density that
influences cosmic acceleration, with its pressure proportional to
energy density within a specific range ($-1$ to $-1/3$). The astonishing
quintessence's boundary case of $-1$ explains the cosmological constant term \cite{Banks:1998vs,Hellerman:2001yi}.
With the acceleration factor provided through the quintessence, the
outer as well as inner horizons of the BH have been thoroughly
examined in Ref.~\cite{Hellerman:2001yi}. Kiselev's work on solving the Einstein
equations for BHs surrounded by quintessence explores its
significant astrophysical effects \cite{Kiselev:2002dx}.

In this section, we briefly review the RN-AdS BH SSCQ and the procedure to obtain its
solution from Einstein's equation. Let us also mention that the solution corresponding to a BH SSCQ has been recently
discovered  \cite{Toledo:2019amt, Chabab:2020ejk, Ma:2019pya, Liang:2020hjz}. The source of such a solution can be considered to be comprised of two distinct sources, one being the quintessence and the other being the cloud of string. Also, these two sources are linearly superposed, meaning no interaction between these two distinct parts. The energy-momentum tensor for such a scenario can be given as follows.
\begin{eqnarray}\label{1R}\quad
T_{t}^{t}=T_{r}^{r}=\rho_{q}+\frac{1}{r^{2}}~,~
T_{\theta}^{\theta}=T_{\phi}^{\phi}=-\frac{\rho_{q}}{2} \ (1+3\omega_{q}) ~, 
\end{eqnarray}
where $\rho_{q}$ implies the density of the quintessence and $\omega_{q}$ represents the quintessential state parameter. Furthermore, equation of state can be given as $p_{q} = \rho_{q} \ \omega_{q}$, where the quintessential pressure ($p_{q}$) correlates with the density and the state parameter  Also, one needs to note that in Eq.~\eqref{1R}, we have considered the spherical symmetry in the spacetime. Therefore, considering the spherical symmetry ansatz in the spacetime, the line element can be given as: 
\begin{eqnarray}\label{2R}
ds^{2}=f(r)dt^{2}-(f(r))^{-1}dr^{2}-(d\theta^{2}+\sin^{2}{\theta}
d\phi^{2})r^{2}. \label{LINEEL}
\end{eqnarray}
 Plugging Eqs.~\eqref{1R} and \eqref{2R} in Einstein's equation (in Anti-de-Sitter (AdS) space), one can obtain the metric parameter $f(r)$ as
\begin{eqnarray}\label{4R}
f(r)=-a-\frac{2 M}{r}+\frac{Q^2}{r^2}-\frac{\alpha }{r^{3 \omega_{q}
+1}}-\frac{\Lambda  r^2}{3}+1~.
\end{eqnarray}
The above solution indicates the presence of a BH in the spacetime, and its horizon is located at $r_h$, where $r_h$ is determined by the condition $f(r_h)=0$. Furthermore, $M$ and $Q$ denote the mass and the charge of the BH that exists in this spacetime. In addition, the cosmological constant of the AdS spacetime is denoted by $\Lambda$, which is related to the AdS curvature ($l$) as $\Lambda=~-3/l^2$,
whereas the integration constant $a$ is the string cloud parameter.
Physically, it demonstrates an extra gravitational effect produced by the string cloud surrounding the BHs. On the contrary, $\alpha$ is the normalization factor associated with the density of
quintessence, and the state parameter
satisfying the condition $-1 \leq~ \omega_{q} < ~-1/3$ are related as
\begin{eqnarray}\label{5R_1}
\rho_{q}=-3\alpha\omega_{q}~/~2r^{3(1+\omega_{q})}.
\end{eqnarray} 
Since our main emphasis is on the study of BH thermodynamics, we replace the cosmological constant ($\Lambda$) with the thermodynamic pressure of the extended phase space formalism \cite{Kastor:2009wy}, which is given as $P=-\frac{\Lambda}{8 \ \pi}$. Thereby, the metric component $f(r)$ looks as:
\begin{eqnarray}\label{5R}
f(r)=-a+\frac{8}{3} \pi  P r^2-\frac{2
M}{r}+\frac{Q^2}{r^2}+1-\frac{\alpha }{r^{3 \omega_{q} +1}}.
\end{eqnarray}
 In the subsequent analysis, we have realized the impact of the state parameter in the presence of the Barrow entropy by fixing the values of $P$, $Q$, $\alpha$, and $a$ while giving variation to the state parameter.

\section{Thermodynamics of Black Hole SSCQ}\label{sec-3} 
This section concentrates on studying the thermodynamics of RN-AdS BH SSCQ through \bre. In BH's thermodynamics, BH's mass is one of the basic properties due to its role in obtaining the thermodynamic behavior, characteristics, and its interaction with its surroundings. Usually, we compute the BH’s mass by employing the following condition: $f(r_{h})=0$, which yields 
\begin{eqnarray}\label{6R}
M=\frac{r_{h}^{-3 \omega_{q} -1}}{6}  \left(-3 a r^{3 \omega_{q} +2}_{h}+8 \pi  P
r_{h}^{3 \omega_{q} +4}+3 Q^2 r_{h}^{3 \omega_{q} }+3 r_{h}^{3 \omega_{q} +2}-3 \alpha
r_{h}\right).
\end{eqnarray}

Since it is known that BHs have intricate and fractal structures at a small scale due to the quantum gravitational effect, which potentially raises their surface area and entropy \cite{Barrow:2020tzx}. Therefore, this intricate horizon structure recommends that BH's entropy might be greater than the standard entropy, which implies that the usual principles for BHs do not remain unchanged. In the quantum field, quantum particle paths are non-differentiable at a small scale, meaning they are not smooth or continuous. In contrast, Bekenstein-Hawking entropy is related to the smooth surface area in which much information might be lost because it doesn't consider the intricate structure of the horizon surface at the quantum scale. In addition, the presence of string cloud and quintessence field can also influence the spacetime geometry, which creates ambiguity in the interpretation of standard entropy and adds complexity to the horizon structure at the small scale. Thereby, by assuming that the surface of the BH horizon has a fractal structure, Ref.~\cite{Barrow:2020tzx} suggests a change in the relation between entropy and the area of BH. When the horizon surface changes proportional to the horizon radius $r_{h}^{\Delta+2}$, it indicates that surface area grows more rapidly than the standard entropy. Thus, the modified entropy takes the following form
\begin{eqnarray}\label{7R}
S_{B}=\bigg(\frac{A}{A_{0}}\bigg)^{1+\frac{\Delta}{2}},
\end{eqnarray}
where the Planck area is represented by $A_{0}$ and the conventional
horizon area by $A$. 
The exponent $\Delta$ (\brp~or fractal correction parameter) shows the quantum gravitational perturbation, with specific values ranging from 0 to 1. When  $\Delta = 0$, the horizon configuration is the simplest, yielding the familiar Bekenstein-Hawking entropy. In contrast, $\Delta = 1$ denotes the horizon's most fractal and complex structure. 
By doing some manipulation, we will obtain the following form of the
entropy
\begin{eqnarray}\label{8R}
S_{B}=\big(\pi r_{h}^{2}\big)^{1+\frac{\Delta}{2}}.
\end{eqnarray}

Its relationship to the horizon radius is monotonic, allowing the BH mass parameter to modify in terms of \bre, charge, and thermodynamic pressure. As we know, in the presence of a cosmological constant, BH's mass cannot be treated as the internal energy of the BH. Instead of internal energy, it should be treated as the enthalpy of the BH
$M(S_{B},~Q,~P)\equiv~H(S_{B},~Q,~P)$ as described in Ref.~\cite{Kastor:2009wy}. In addition, thermodynamic pressure in the presence of a cosmological constant follows from the relation $P=-\Lambda/8\pi G$. Thereby, our mass in terms of \bre~ takes the given form as
\begin{eqnarray}\label{9R}
M(S_{B}, Q, P)=\frac{ S_{B}^{\frac{1}{\Delta
+2}}}{2\sqrt{\pi}}-\frac{a S_{B}^{\frac{1}{\Delta
+2}}}{2\sqrt{\pi}} + \frac{4 P S_{B}^{\frac{3}{\Delta
+2}}}{3\sqrt{\pi}}+\frac{Q^{2}\sqrt{\pi}}{2S_{B}^{\frac{1}{\Delta
+2}}}-\frac{\alpha \ \pi^{3\omega_{q}/2}}{2 S_{B}^{\frac{3
\omega_{q}}{\Delta +2}}}.
\end{eqnarray}
 
Furthermore, the temperature is another significant thermodynamic property in BH thermodynamics, which is crucial for BH's stability and phase transition. The temperature for the RN-AdS BH SSCQ in terms of \bre~is derived using the given relation
\begin{eqnarray}\nonumber
T_{B}&=&\left(\frac{\partial{M}}{\partial{S_{B}}}\right)_{P,~Q,~a,~\alpha}\\\label{11R}&=&\frac{1}{\Delta
+2}\left(-\frac{a}{2\sqrt{\pi} S_{B}^{\frac{\Delta +1}{\Delta
+2}}}+\frac{4PS_{B}^{\frac{1-\Delta}{\Delta +2}}}{\sqrt{\pi}
}-\frac{Q^{2}}{2 S_{B}^{\frac{\Delta +3}{\Delta +2}}}+\frac{3
\alpha \omega_{q}\pi^{\frac{3\omega_{q}}{2}}}{2
S_{B}^{\frac{3\omega_{q}}{\Delta
+2}+1}}+\frac{1}{2\sqrt{\pi}S_{B}^{\frac{\Delta+1}{\Delta+2}}}\right).
\end{eqnarray}

\begin{figure} \centering
\epsfig{file=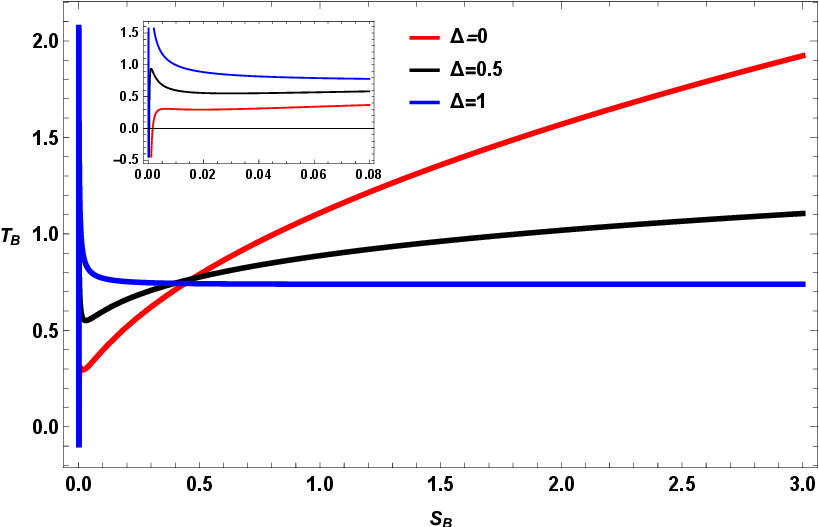,width=.45\linewidth} \caption{\raggedright $T_{B}$
versus $S_{B}$ in the presence of \bre~by putting $\Delta=0$ Hawking-Bekenstein case (red curve), $\Delta=0.5$ (black curve) and $\Delta=1$ most intricate structure (blue curve).} \label{fig-2}
\end{figure}

The behavior of $T_{B}$ is graphically obtained by setting $\alpha=0.3,~a=0.8,~P=1,~Q=0.01$ for $\Delta=0$ (red curve), $\Delta=0.5$ (black curve) and $\Delta=1$ (blue curve) in 
Fig.~\ref{fig-2}.  We observe that the temperature $T_{B}$ initially increases with negative behavior, but after some ranges of $S_{B}$, it becomes positive. This temperature $T_{B}$ transition from negative to positive behavior indicates a phase transition. In the right panel, we observe that the temperature behavior does not increase rapidly as the fractal correction parameter increases. Similarly, one can easily compute the thermodynamic volume $(V)$ and potential $(\Phi)$ by employing Eq.~\eqref{9R}, which yields 
\begin{eqnarray}\label{12R}
\Phi=\left(\frac{\partial{M}}{\partial{Q}}\right)_{P,~S_{B},~a,~\alpha}=\frac{Q\sqrt{\pi}}{S_{B}^{\frac{1}{\Delta+2}}}=Q/r_{h}\
, \quad V=\left(\frac{\partial{M}}{\partial{P}}\right)_{S_{B},~Q,~a,~\alpha}=
\frac{4S_{B}^{\frac{3}{\Delta+2}}} {3\pi^{3/2}}=\frac{4}{3}\pi
r_{h}^{3}~.
\end{eqnarray}

\begin{figure} \centering
\epsfig{file=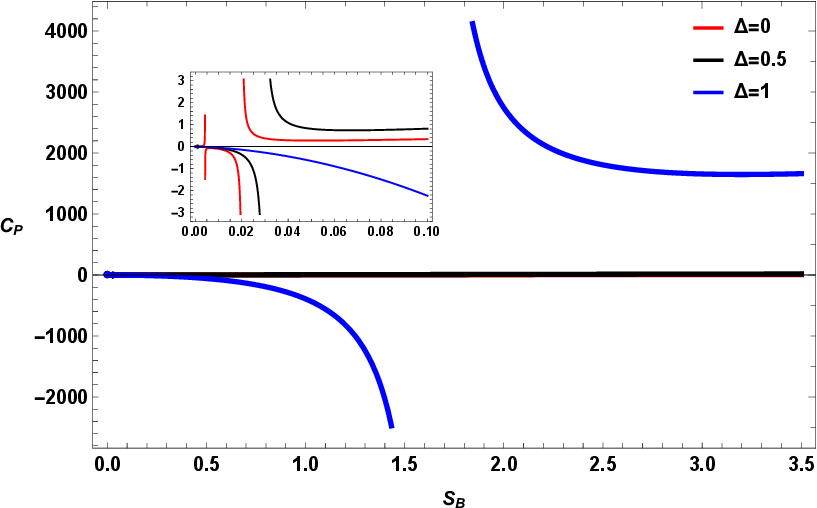,width=.45\linewidth} \caption{\raggedright $C_{P}$
versus $S_{B}$ in the presence of \bre~by inserting $\Delta=0$ (red curve), $\Delta=0.5$ (black curve) and $\Delta=1$ (blue curve).} \label{fig-3}
\end{figure}

The thermodynamic volume $V$ increases monotonically with the horizon radius and the
electric potential $(\Phi)$ corresponds to the classical electromagnetism
model. We obtain the internal energy $U(S_{B}, Q, V)$ for RN-AdS BH SSCQ from the enthalpy relation in Eq.~\eqref{9R} by utilizing the Legendre transformation, and its expression is given as   
\begin{eqnarray}\label{13R}
U(S_{B},~Q,~V)=M(S_{B},~Q,~P)-PV~,
\end{eqnarray}
\begin{eqnarray}\label{14R}
U(S_{B},~Q,~V)=\frac{1}{2}
\left((1-a)\frac{S_{B}^{\frac{1}{\Delta+2}}}{\sqrt{\pi}}+\frac{Q^{2}
\sqrt{\pi}}{S_{B}^{\frac{1}{\Delta+2}}} -\frac{\alpha \
\pi^{\frac{3\omega_{q}}{2}}}{S_{B}^{\frac{3\omega_{q}}{\Delta+2}}} \right)~.
\end{eqnarray}
Heat capacity at constant pressure $(C_{P})$ is considered as one of the methods to analyze BH's stability. If its behavior is positive, the BH is stable; if it shows negative behavior, the BH is unstable. Furthermore, we aim to observe the Davies type phase-transition from the heat capacity's behavior by employing Eq.~\eqref{11R}, which takes the following form as discussed in Refs.~\cite{Davies:1989ey, Lousto:1994jd,Muniain:1995ih, Bhattacharya:2024bjp}. By employing Eq.~\eqref{11R}, one can obtain the heat capacity equation concerning \bre, which turns out
\begin{eqnarray}\label{CDTP}
C_{\mathrm{Y}_{\mathrm{i}}}=T\left(\frac{\partial S}{\partial T}\right)_{\mathrm{Y}_{\mathrm{i}}}.
\end{eqnarray}
By utilizing the following argument 
\begin{eqnarray}\label{CDTP1}
\frac{\partial }{\partial S_{B}}\left(\frac{1}{T_{B}}\right)=-\frac{1}{T^{2}}\left(\frac{\partial S}{\partial T}\right)_{\mathrm{Y}_{\mathrm{i}}},
\end{eqnarray}
it is straightforward to obtain
\begin{eqnarray}\label{CDTP2_1}
C_{\mathrm{Y}_{\mathrm{i}}}=-\frac{1}{\frac{1}{T_{B}}\frac{\partial }{\partial S}\left(\frac{1}{T_{B}}\right)_{\mathrm{Y}_{\mathrm{i}}}}~.
\end{eqnarray}
The fact that $T \neq 0$ leads to the following relation 
\begin{eqnarray}\label{CDTP2}
\frac{\partial }{\partial S}\left(\frac{1}{T_{B}}\right)_{\mathrm{Y}_{\mathrm{i}}}=0~.
\end{eqnarray}
By employing Eq.~\eqref{11R}, one can obtain the heat capacity equation concerning \bre, which gives
\begin{eqnarray}\label{15R}
C_{P}=\frac{(\Delta +2) S \left(-a \mathcal{L}+8 \pi  P \mathcal{L}^3-\frac{Q^2}{\mathcal{L}}+3 \alpha  \omega  \mathcal{L}^{-3 \omega }+\mathcal{L}\right)}{-a (\Delta +1) \mathcal{L}+8 \pi  (\Delta -1) P \mathcal{L}^3+\frac{(\Delta +3) Q^2}{\mathcal{L}}+3 \alpha  \omega  (\Delta +3 \omega +2) \mathcal{L}^{-3 \omega }+\Delta  \mathcal{L}-\mathcal{L}},
\end{eqnarray}

where $\mathcal{L}=\left(\pi ^{-\frac{\Delta }{2}-1}
S_{B}\right)^{\frac{1}{\Delta +2}}$. It is quite easy to determine the phase transition or divergence point from Eq.~\eqref{15R} by putting the denominator of heat capacity equal to zero. Here, we mention that we have determined the point of divergence of $C_{P}$ by putting $\omega_{q}=-2/3,~P=1,~Q=0.01,~a=0.8,~\alpha=0.3$ and $\Delta=0,~0.5,~1$. We obtain a divergence point $S_{B_{0}}=0.0187$ for $\Delta=0$ and also for $\Delta=0.5$, we determine two divergence point at $S_{B_{0.5}}=0.029728$. Similarly, we compute a divergence point $S_{B_{1}}=1.63871$ for $\Delta=1$. Furthermore, heat capacity at constant volume can be written as 
\begin{eqnarray}\label{155R}
C_{V}=~\bigg(\frac{\partial{S_{B}}}{\partial{T_{B}}}\bigg)_{V}T=~0~.
\end{eqnarray}
It is easy to verify from Eq.~\eqref{12R} that $C_{V}$ is zero, owing to the link between the entropy and volume \cite{Toledo:2019amt}. In our case, the relation between entropy and volume is $S_{B}\sim V^{\frac{2+\Delta}{3}}$, which indicates that if $V$ is fixed or (constant), then the entropy $S_{B}$ is also fixed. n Fig.~\ref{fig-3}, we illustrate the behavior of $C_{P}$ in terms of \bre by setting $\alpha=0.3,~a=0.8,~P=1,~\omega=-2/3$ and $Q=0.01$.
for the various values of  $\Delta$. The behavior of the heat capacity graph is both negative and positive, indicating instability and stability of the BH in Fig.~\ref{fig-3}. The transition of heat capacity from unstable to stable region is for all the values of  $\Delta$, which also validates the effect of the fractal correction parameter. 

\begin{figure} \centering
\epsfig{file=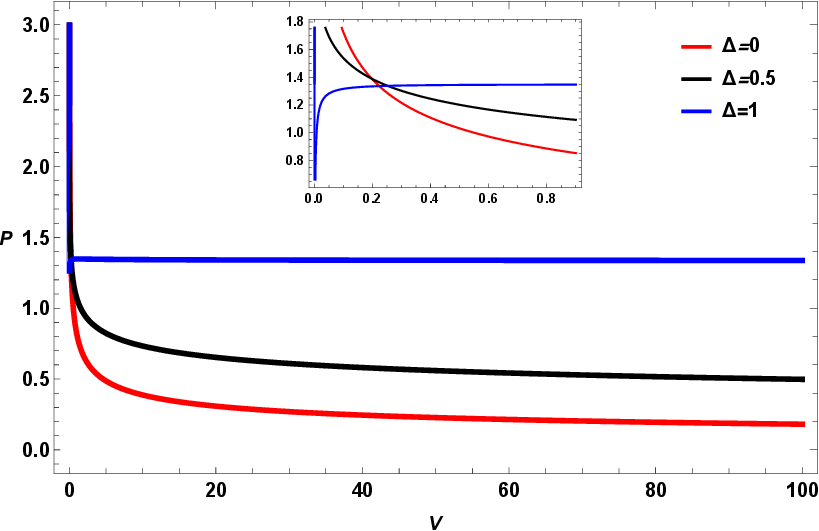,width=.45\linewidth} \caption{\raggedright $P$
in terms of $V$ in the presence of \bre by putting $\Delta=0$ (red curve), $\Delta=0.5$ (black curve) and $\Delta=1$ (blue curve).} \label{fig-4}
\end{figure}

In BH thermodynamics, we also analyze BH stability by studying the $P-V$ phase diagram. Thereby, we discuss the $P-V$ diagram for the RN AdS BH SSCQ in terms of \bre~  by using Eqs.~\eqref{11R} and \eqref{12R}, which we write as 
\begin{eqnarray}\label{16R}
T_{B}(P,~V)=\left(\frac{4}{3\ \sqrt{\pi}
V}\right)^{\frac{\Delta+2}{3}}\left(\frac{Q^{2}}{2(2+\Delta)\chi}+\frac{\chi(1-a)}{2(2+\Delta)}+
\frac{4\pi P \chi^{3}}{2+\Delta}+\frac{3\alpha
\omega_{q}}{2(2+\Delta)\chi^{3\omega_{q}}}\right),
\end{eqnarray}
where $\chi=~\left(\frac{3 V}{4\pi}\right)^{1/3}$. The pictorial behavior of the $P-V$ diagram depicts the variation of
$\Delta=0$ (red curve), $\Delta=0.5$ (black curve) and $\Delta=1$ (blue curve) by using $\alpha=0.3,~a=0.8$ and $Q=0.01$. In Fig.~\ref{fig-4}, one can notice that pressure decreases as the volume increases, which predicts the BH's stability for $\Delta=0,~0.5$ (red and black curve, respectively). But in this figure, as the intricacy of the horizon increases $\Delta=1$ (blue curve), we observe that the pressure is decreasing and, after reaching its minimum point, it rises again as the volume increases, which indicates the impact of the fractal correction parameter.

Furthermore, if we consider that string cloud parameter $(a)$ and normalization factor ($\alpha$) are extensive parameters, then the first law of BH thermodynamics takes shape as discussed in Ref.~\cite{Toledo:2019amt}. The general expression for deriving the first law of BH thermodynamics is obtained directly from the Einstein field equations, as given in Ref.~\cite{Bhattacharya:2021lgk}.
\begin{eqnarray}\label{GFLT}
\delta{M}=T \delta{S}-\frac{r_{h}}{2}\frac{\partial{f(r,M,x^{i})}}{\partial{x^{i}}}dx^{i},
\end{eqnarray}
where $x^{i}=(x^{1},~x^{2},~x^{3},~x^{4})= (P,~Q,~a,~\alpha)$ in our case. We also obtain the first law of BH thermodynamics for RN-AdS BH SSCQ utilizing the generalized formulation for spherically symmetric spacetime, which agrees with the aforementioned first law of BH thermodynamics and yields
\begin{eqnarray}\label{17R}
\delta{M}=T_{B} \delta{S_{B}}+\Phi\ \delta{Q}+V \delta{P}+ \mathcal{C}
\ \delta{a}+\mathcal{D} \ \delta{\alpha}~,
\end{eqnarray}
where $\mathcal{C}$ and $\mathcal{D}$ are the conjugate to the
parameters $a$ and $\alpha$

\begin{eqnarray}\label{18R}
\mathcal{C}=-\frac{S_{B}^{\frac{1}{\Delta +2}}}{2 \ \sqrt{\pi }}\ ,
\quad
\mathcal{D}=-\frac{1}{2}\left(\frac{\sqrt{\pi}}{S_{B}^{\frac{1}{\Delta
+2}}}\right)^{3\ \omega_{q}}.
\end{eqnarray}

Similarly, by adopting the methodology described in Ref.~\cite{Bhattacharya:2021lgk},  we derive the generalized Smarr formula for spherically symmetric spacetime from the Einstein field equations, which is given as
\begin{eqnarray}\label{SmrR}
M=2TS-\frac{r^{2}_{h}}{2}\left(\frac{\partial{f(r,~M,~Q,~a,~\alpha)}}{\partial{r}}\right)|_{r=r_{h},M=0}~.
\end{eqnarray}
Before moving directly to the Smarr relation for RN-AdS BH SSCQ, we want to mention that $T$ is the conventional Hawking temperature, and $S$ is the Bekenstein-Hawking entropy. As we discuss BH thermodynamics by employing \bre, some modifications are required, as given in Ref.~\cite{Ladghami:2024sen}.
\begin{eqnarray}\label{MSMR}
TS=\frac{(\Delta+2)}{2}T_{B}S_{B}~.
\end{eqnarray}
Thereby,  by plugging Eq.~\eqref{MSMR} into Eq.~\eqref{SmrR}, the generalized Smarr formula can be found for any spherically symmetric case in terms of \bre. The generalized Smarr formula for any spherical symmetric spacetime in the context of \bre~takes the following shape
\begin{eqnarray}\label{GSmrR}
M=(\Delta+2)T_{B}S_{B}-\frac{r^{2}_{h}}{2}\left(\frac{\partial{f(r,~M,~Q,~a,~\alpha)}}{\partial{r}}\right)|_{r=r_{h},M=0}~.
\end{eqnarray}
Moreover, the distinct feature of our approach is that by putting $\Delta=0$, we can retrieve the generalized Smarr formula for the Bekenstein-Hawking entropy case, which applies to any spherically symmetric spacetime. Therefore, the Smarr relation in our case, which is RN-AdS BH SSCQ, is obtained
by inserting Eq.~\eqref{5R} into Eq.~\eqref{GSmrR}. Hecne, the Smarr relation can be represented as
\begin{eqnarray}\label{SmrrR}
M=(2+\Delta)T_{B}S_{B}-2PV+\Phi Q+ (3\omega_{q}+1)\mathcal{D} \ \alpha~.
\end{eqnarray}
Furthermore, the Smarr relation for RN-AdS BH SSCQ, which we obtain from the generalized formula of spherical symmetric spacetime, also agrees with the one given in Ref.~\cite{Chabab:2020ejk}.

\section{Thermodynamic Topology of Black Holes SSCQ in Terms of \bre}\label{sec-5}
This section focuses on the topological interpretation of the RN-AdS BH SSCQ by employing \bre. The thermodynamic topology observes the phase transition or divergence in the behavior of RN-AdS BHs SSCQ. It is worth mentioning that many techniques analyze the critical points or phase transition in BH thermodynamics, and we also adopt some methods, but they do not provide any concrete information regarding this. For example, we employ the thermodynamic geometry formalisms, but they do not give us any information regarding the divergence or phase transition of heat capacity due to the complicated thermodynamic potentials. Therefore, we employ the thermodynamic topology because it is a unique approach to discussing the characteristics of BHs without exploring their detailed framework and particular field equations.

Now, we provide a basic framework to discuss the thermodynamic topology for exploring the RN-AdS BH SSCQ phase transition in terms of \bre. Firstly, we discuss Duan's formalism in the context of topological charge, which we employ in the phase space of the thermodynamic space \cite{Duan:1979ucg, Duan:1984ws}. As described in Ref.~\cite{Bhattacharya:2024bjp}, we obtain the thermodynamic potential by employing Eq.~\eqref{CDTP2}, which yields 
\begin{eqnarray}\label{THP}
\Phi(S_{B},~\theta)=~\frac{1}{\sin{\theta} \ T(S)_{Y_{i}}}~.
\end{eqnarray}
We mention here that the additional term $1/\sin \theta$ is incorporated in the thermodynamic potential by using the concept given in Refs.~\cite{Wei:2021vdx, Bhattacharya:2024bjp} which simplifies the analysis of topology. Therefore, its vector field takes the given shape
\begin{eqnarray}\label{THPP1}
\phi^{S_{B}}=\partial_{S_{B}} \Phi(S_{B},~\theta)&=&~\frac{1}{\sin{\theta}} \frac{\partial}{\partial S_{B}} \left(\frac{1}{T_{B}}\right)_{{\mathrm{Y}}_{\mathrm{i}}},\\\label{THPP2}
\phi^{\theta}=\partial_{\theta}\Phi(S_{B},\theta)&=&-\frac{\cot{\theta}\csc{\theta}}{T(S_{B})_{Y_{i}}}~.
\end{eqnarray}
If we see Eqs.~\eqref{THPP1} and \eqref{THPP2}, we realize the significance of the expression $1/\sin {\theta}$ because of various reasons. For example, this additional term assists in determining the divergence point or critical point in the $S-\theta$ plane. If we put $\theta=\pi/2$ then $\phi^{\theta}$ vanishes which indicates that the zero point of $\phi$ lies at $\theta=\pi/2$. Furthermore, if we put $\theta=0,~\pi$ then it is perpendicular to vector $\phi$ which we assumed to be the boundary of the parameter space. Now, according to Duan's mapping theory and by using the anti-symmetrical superpotential $V^{\mu\nu}=1/2\pi (\epsilon^{\mu\nu\rho} \epsilon_{ue}n^{u}\partial_{\rho}n^{e})$, it is quite easy to derive the topological current $j^{\mu}$ as discussed in Refs.~\cite{Wei:2021vdx, Bhattacharya:2024bjp}.
\begin{eqnarray}\label{THP1}
j^{\mu}=\partial_{\nu} V^{\mu\nu}=1/2\pi \  \epsilon^{\mu\nu\rho} \ \epsilon_{ue} \ \partial_{\nu}n^{u}\partial_{\rho}n^{e},
\end{eqnarray}
where $\mu,~\nu,~\rho=~0,~1,~2$,  the normalized vector is presented by $n^{u}$ (which satisfies  $n^{u}n_{u}=1$ and $n^{u}\partial_{\nu}n^{u}=0$ )  and $j^{\mu}$  is the topological current. As we mentioned earlier, $V^{\mu\nu}$ is anti-symmetric, and by using this characteristic, it is quite easy to obtain that $j^{\mu}$ is conserved.
\begin{eqnarray}\label{THP1}
\partial_{\mu}j^{\mu}=0~.
\end{eqnarray}
After making some adjustments as discussed in Ref.~\cite{Bhattacharya:2024bjp}, one can get the topological current, which is further simplified by using the Jacobi tensor; it takes the given shape 
\begin{eqnarray}\label{JBT}
\epsilon^{ce} J^{\mu} (\frac{\phi}{y})=\epsilon^{\mu\nu\rho}\partial_{\nu} \phi^{c} \partial_{\nu} \phi^{c} \partial_{rho} \phi^{b},
\end{eqnarray}
where Jacobian is presented by $J(\phi/y)=\frac{\partial \left(\phi^{1},~\phi^{2}\right)}{\partial \left(y^{1},~y^{2}\right)}$. So, the term $j^{\mu}$ is computed in the form of a Jacobian tensor as given in  
\begin{eqnarray}\label{JBT1}
j^{\mu}=\frac{1}{2\ \pi} \left(\gamma_{\phi}\ln{||\phi||}\right)J^{\mu}\left(\frac{\phi}{y}\right),
\end{eqnarray}
where $\gamma_{\phi}=\frac{\partial}{\partial \ \phi^{u}}\ \frac{\partial}{\partial \ \phi^{u}}$. So, the final expression of the current follows from the Laplacian Green function of 
$\phi$ space, which is a fundamental result in two-dimensional potential theory. According to this Laplacian of the logarithm of the magnitude of $||\phi||$ is proportional to the two-dimensional delta function centered at $\phi$ equal to zero, and it can be written as
\begin{eqnarray}\label{JBT}
j^{\mu}=\delta^{2}(\phi)J^{\mu}\left(\frac{\phi}{y}\right),
\end{eqnarray}
and One can easily derive the total charge inside $\mathcal{C}$ which takes the given shape
\begin{eqnarray}\label{JBT1}
Q=\int_{C}\delta^{2}(\phi)J^{\mu}\left(\frac{\phi}{y}\right)d^{2}y~.
\end{eqnarray}
Thereby, from the equations above, we observe that at $\phi=0$, the topological current is non-vanishing, and similarly, $\mathcal{C}$ encloses the zero point of $\phi$ $iff$ the topological charge is non-vanishing. There will be no charge if no zero point is enclosed in $\mathcal{C}$. By doing some manipulations, total topological charge takes the given form  \begin{eqnarray}\label{JBT2}
Q=\sum^{N}_{i=1}\beta_{i} \ \eta_{i}=\sum^{N}_{i=1}\omega_{i}~,
\end{eqnarray}
where $\beta_{i}$ is the non-negative integer number known as Hopf index, which appears when $y^{\mu}$ vector revolves around the zero point $x_{i}$ in the $\phi$ space and $\eta_{i}$ is the Brower degree which is equal to the $\text{sign}\left(J^{0}(\frac{\phi}{y})\right)_{x_{i}}=\pm 1$. Furthermore, $\omega_{i}$ is the winding number corresponding to the $i$-th zero point enclosed in $\mathcal{C}$. Moreover, if one assumes that $\mathcal{C}_{i}$ is a positive closed curve that captures the $i$-th zero point 
then corresponding to that $i$-th zero point, one can easily get the winding number, which is written as 
 \begin{eqnarray}\label{JBT33}
\omega_{i}=\frac{1}{2 \ \pi} \oint_{\mathcal{C}_{i}}d\Omega~.
\end{eqnarray}
Here, $\Omega$ epresents the deflection angle of the vector field as it follows the given contour. Mathematically, it is expressed as $\Omega=\arctan{\left(\phi^{2}/ \Phi^{1}\right)}$. So, we obtain $d\Omega$ by employing the following relation $n^{1}dn^{2}-n^{2}dn^{1}$, where $n^{1}$ and $n^{2}$ are the unit vectors. It is worth mentioning here that we discuss the thermodynamic topology of the RN-AdS BH SSCQ by substituting $\omega=-2/3$ for simplicity. By employing the above formalism, we explore the thermodynamic topology to identify the divergence or phase transition point in RN-AdS BH SSCQ. Firstly, with Eq.~\eqref{THP}, we define our thermodynamic potential as
\begin{eqnarray}\label{RNJBT}
\Phi(S_{B},\theta)=\frac{1}{\sin{\theta}}\frac{2 (\Delta +2) S_{B}}{-a \mathcal{L}+8 \pi  P \mathcal{L}^{3}-Q^2 \mathcal{L}^{-1}-2 \alpha  \mathcal{L}^{2}+\mathcal{L}}~.
\end{eqnarray}

The vector field components are obtained by utilizing Eqs.~\eqref{THPP1},~\eqref{THPP2}, and \eqref{RNJBT}, which takes the given form
\begin{eqnarray}\label{RNJBT1}
&\phi^{S_{B}}&=\frac{2 \mathcal{L} \bigg[\mathcal{L}^{2} \bigg\{-a (\Delta +1)+\Delta +8 \pi  (\Delta -1) P \mathcal{L}^{2}-2 \alpha  \Delta  \mathcal{L}+1\bigg\}-(\Delta +3) Q^2\bigg]}{\sin (\theta )\bigg[\mathcal{L}^{2} \bigg\{a+2 \mathcal{L} \left(\alpha -4 \pi  P \mathcal{L}\right)-1\bigg\}+Q^2\bigg]^2}~,\\\label{RNJBT2}
&\phi^{\theta}&=-\frac{2 (\Delta +2) S \cot (\theta ) \csc (\theta ) \mathcal{L}}{\mathcal{L}^{2} \bigg[-a+8 \pi  P \mathcal{L}^{2}-2 \alpha  \mathcal{L}+1\bigg]-Q^2}~.
\end{eqnarray}

\begin{figure}
     \centering
     \begin{subfigure}[b]{0.45\textwidth}
         \centering
         \includegraphics[width=\textwidth]{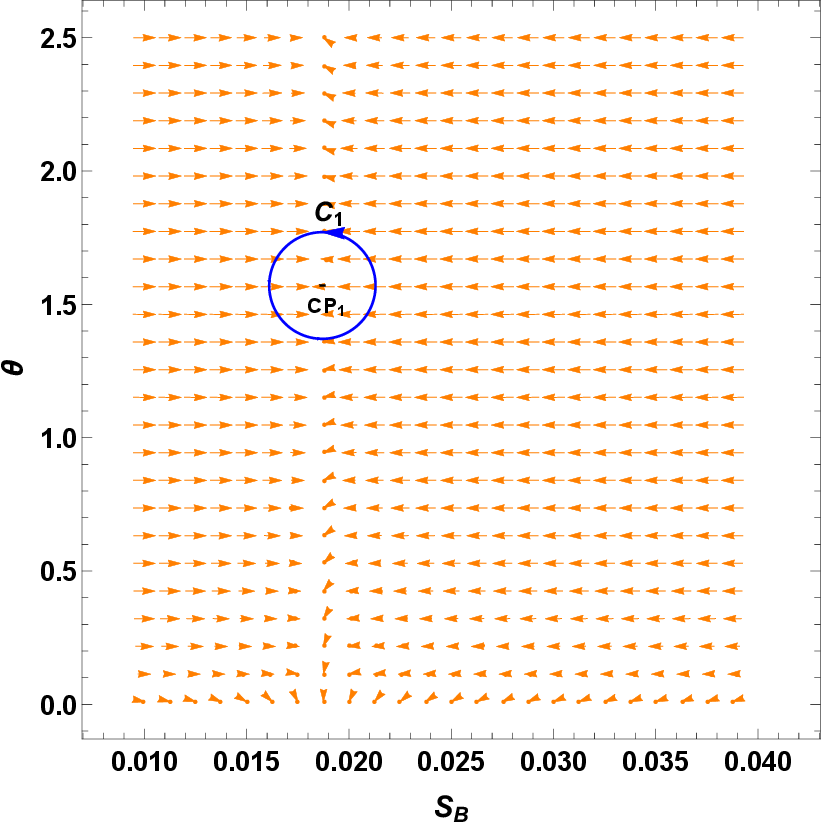}
         \caption{$\Delta=0$}
         \label{TT1}
     \end{subfigure}
      \begin{subfigure}[b]{0.45\textwidth}
         \centering
         \includegraphics[width=\textwidth]{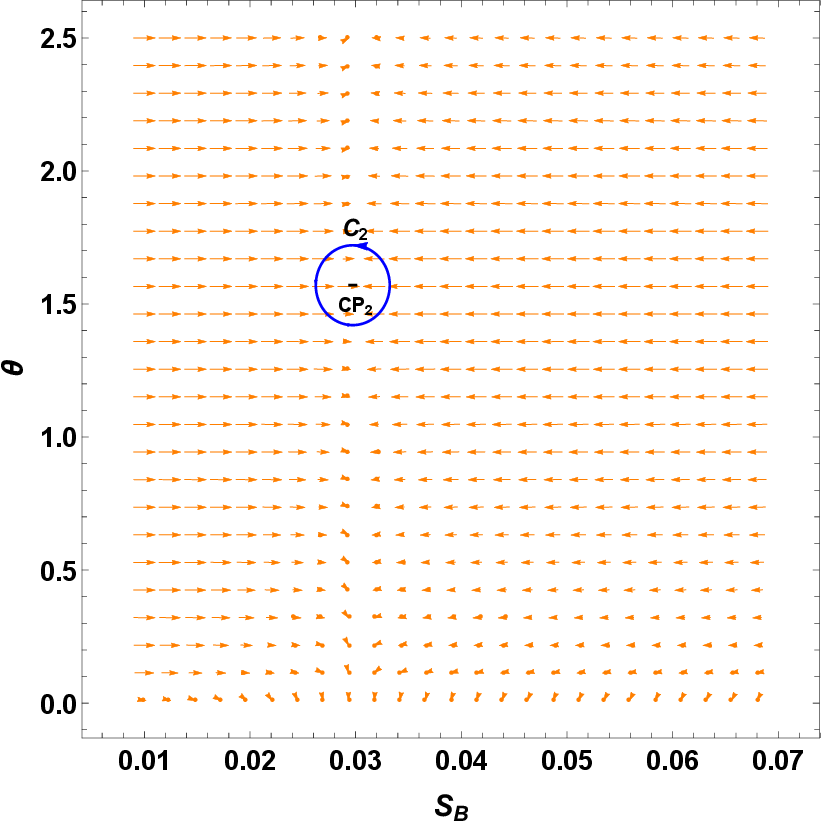}
         \caption{$\Delta=0.5$}
         \label{TT2}
     \end{subfigure}
     \begin{subfigure}[b]{0.45\textwidth}
         \centering
         \includegraphics[width=\textwidth]{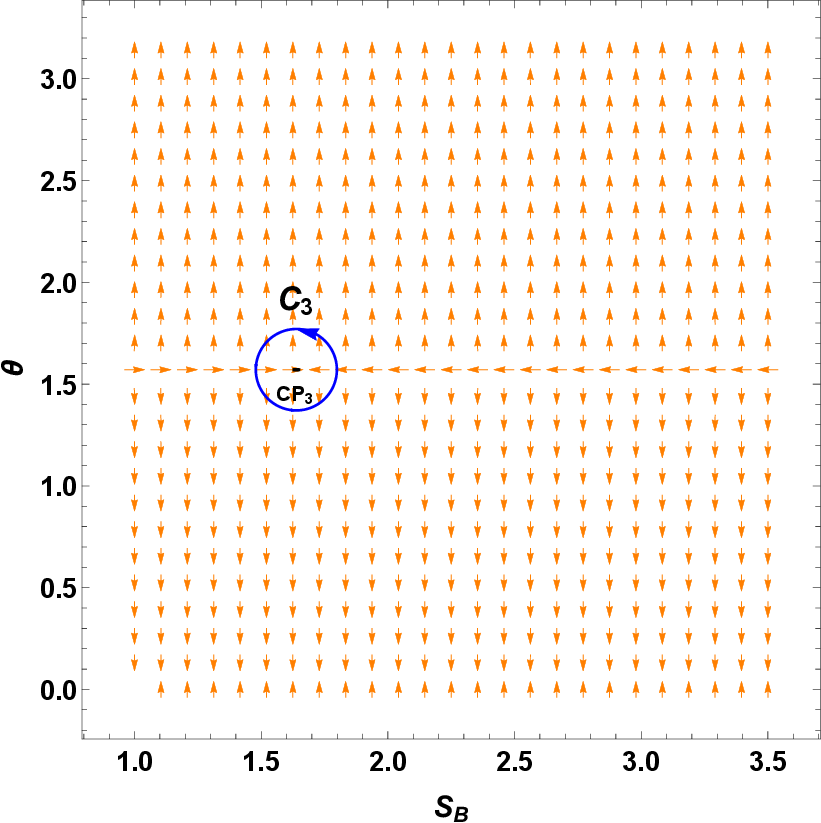}
         \caption{$\Delta=1$}
         \label{TT3}
     \end{subfigure}
        \caption{The orange arrow depicts the normalized vector in the $S_{B}-\theta$ plane. The black dot enclosed in the blue contour denotes the critical point or divergence point for $\Delta=0$ (smooth surface referred to as Hawking Bekenstein entropy case) in Fig.~\subref{TT1}, $\Delta=0.5$ and $\Delta=1$ (Most intricate horizon structure) in Fig.~\subref{TT2} and \subref{TT3}, respectively.}
        \label{TTT}
\end{figure}
In Fig.~\ref{TTT}, we present the unit vector $n$ in $S_{B}$ versus $\theta$ plane by using the following expressions to derive the normalized vector $n^{u}=\left(\frac{\phi^{s}}{\|\phi\|},~\frac{\phi^{\theta}}{\|\phi\|}\right)$. It observes that we have zero points or critical points at $CP_{1},~CP_{2}$ and $CP_{3}$. We assume a loop of the circle $C_{1}$ that captures the zero points or divergence points for $\Delta=0$ at $CP_{1}$ in Fig.~\ref{TTT}(\subref{TT1}). Similarly, we encircle the zero points $CP_{3}$  by putting $\Delta=0.5$ in Fig.~\ref{TTT}(\subref{TT2}),  while we enclose the zero point $CP_{3}$ by $C_{3}$ for $\Delta=1$ in Fig.~\ref{TTT}(\subref{TT3}). Let us mention here that the topological charge or winding number does not depend on the circles which enclose the zero points. Hence, we define a contour that is parameterized  by $\psi$ for Fig.~\ref{TTT}, and it is given as
\begin{eqnarray}\label{RNJBT1}
&S_{B}&=u \cos{\psi}+S_{B_{P}}~,\\\label{RNJBT2}
&\theta&=u \sin{\psi}+\pi/2~,
\end{eqnarray}
where $u$ depicts the radius and it is centered at $(S_{B_{P}},~\pi/2)$. For Fig.~\ref{TTT}(\subref{TT1}), we choose $u=0.4$ and $S_{B_{0}}=CP_{1}$, in Fig.~\ref{TTT}(\subref{TT2}), we employ $u=0.1$ and $S_{B_{0.5}}=CP_{2}$ while in case of Fig.~\ref{TTT}(\subref{TT2}), we select $u=0.4$ and $S_{B_{1}}=CP_{3}$. Now, by using Eq.~(\ref{JBT33}), one can compute the deflection angle which is given as 
\begin{eqnarray}\label{RNJBT3}
\Omega(\psi) =\int^{\psi}_{0} 
|\epsilon_{ue}n^{u}\partial_{\psi} n^{e}|d\psi~, 
\end{eqnarray}
and we obtain topological charge for Fig.~\ref{TTT} from Eq.~\eqref{RNJBT3} as follows
\begin{eqnarray}\label{RNJBT4}
Q=\pm \frac{\Omega(2 \ \pi)}{2 \ \pi}~.
\end{eqnarray}
We present the $S_{B}-\theta$ plane for $\Delta=0,~0.5$ and $\Delta=1$ to analyze the impact of the intricacy of the horizon. Here, we mention that for $\Delta=0,~0.5$, we modify Eq.~\eqref{RNJBT3} by applying the condition of the mod because the CPs are very close to zero at the origin, which produces the complex topological charge. Therefore, using this argument, we observe that in the case of $\Delta=0$, the topological charge 
 with respect to the $CP_{1}$ is $Q=0.9650 \approx 1$, respectively while for $\Delta=0.5$, we compute $Q=1.007\approx1$ for $CP_{2}$. Furthermore, in the case of $\Delta=1$, it provides the topological charge $Q=1$.
\section{Conclusions}\label{sec-6}
In this work, we have considered RN-AdS BH SSCQ as a complicated thermodynamic system for studying its thermodynamics and thermodynamical topology via \bre. We
obtained thermodynamic parameters analytically by
assuming the cosmological constant as the thermodynamic pressure and observed their behavior for the quintessence state parameter $\omega_{q}$, and the fractal correction parameter $\Delta$. We have used $\Delta=0$ for the smooth surface, which takes us to the conventional entropy, and $\Delta=1$ for the most intricate (complex) structure, to graphically witness the impact of fractal correction parameter on the thermodynamics of RN-AdS BH SSCQ. It can be observed that as the fractal correction parameter increases, the horizon structure becomes more and more intricate, influencing the thermodynamic behavior of the RN-AdS BH SSCQ. For example, in the $P-V$ graph, it can be observed that when the fractal correction parameter approaches its most intricate values, its graph depicts the decreasing to increasing behavior rather than smooth decreasing behavior, which is significant because it is associated with the phase transition. Also, in the heat capacity graph, we can see the impact of the fractal correction parameter by suppressing the phase transition at $\Delta=1$. Furthermore, the additional field (string cloud and quintessence fields) adds complexity, which also influences the thermodynamic behavior of these quantities. Our results signify that the quantum gravitational effects modified the thermodynamic behavior of the BH due to the intricate horizon structure. Let us mention here that the string cloud parameter $(a)$ and normalization factor $(\alpha)$ play an important role in the stability of this BH. In addition, we have developed the first law of BH thermodynamics and the Smarr relation in terms of \bre~ for RN-AdS BH SSCQ by employing the generalized formula for spherical symmetric spacetime computed from the Einstein field equation. We have modified the generalized formula for the Smarr relation \cite{Bhattacharya:2021lgk} by doing some corrections to make it consistent for any spherical symmetric spacetime in terms of \bre~and our computed generalized Smarr formula in the context of \bre~ is applicable and consistent for any spherically symmetric spacetime. Furthermore, the Smarr relation for RN-AdS BH SSCQ in terms of \bre~ is consistent, too.

Furthermore, we have graphically described the topological interpretation of critical points that are observed in the heat capacity behavior for $\Delta=0,~0.5,~1$ at $CP_{1}, CP_{2},~CP_{3}$. However, the non-zero values of topological charge also predicted the existence of critical points.
 In our investigation, we have observed that for a smooth horizon, $\Delta=0$ and  $\Delta=0.5$ provides us with a complex topology. We addressed this problem by using modulus operation $mod$, which gives us a topological charge $Q\approx 1$. In the case of the most intricate horizon structure $\Delta=1$, it gives us the topological charge $Q=1$. Moreover, a topological charge of the BHs depicted their stability and instability. For example, if the topological charge is $1$, it is considered stable; in the case of $-1$, it is unstable. In our analysis, we have observed that for $\Delta=0,~0.5,~1$ RN-AdS BH SSCQ is stable for all the cases of the Barrow entropy. 
 
We aim to extend our analysis to rotating BHs and interpret the phenomenon of phase transition in these type of BHs by using our formalism. For instance, one can examine how Barrow's fractal structure interacts with phase transitions in BH thermodynamics might enhance our understanding of quantum spacetime. Exploring how this framework adapts to rotating BHs and ensuring its consistency with holographic principles could enhance our understanding of the relationship between quantum mechanics and GR.

\section*{Acknowledgments}

This work of K.Bhattacharya is supported partly by the New Faculty Seed Grant (NFSG) of BITS Pilani, Dubai Campus and partly by the JSPS KAKENHI Grant (Number: 23KF0008). Moreover, the research work of K. Bamba is supported by the JSPS KAKENHI Grant (Numbers: 21K03547, 24KF0100)


\begin{thebibliography}{50}
\bibitem{Letelier:1979ej}
P.~S.~Letelier,
Phys. Rev. D \textbf{20}, 1294-1302 (1979).

\bibitem{Chamblin:1999tk}
A.~Chamblin, R.~Emparan, C.~V.~Johnson and R.~C.~Myers,
Phys. Rev. D \textbf{60}, 064018 (1999).

\bibitem{Chamblin:1999hg}
A.~Chamblin, R.~Emparan, C.~V.~Johnson and R.~C.~Myers,
Phys. Rev. D \textbf{60}, 104026 (1999).

\bibitem{Sekiwa:2006qj}
Y.~Sekiwa,
Phys. Rev. D \textbf{73}, 084009 (2006).

\bibitem{Caldarelli:1999xj}
M.~M.~Caldarelli, G.~Cognola and D.~Klemm,
Class. Quant. Grav. \textbf{17}, 399-420 (2000).

\bibitem{Kastor:2009wy}
D.~Kastor, S.~Ray and J.~Traschen,
Class. Quant. Grav. \textbf{26}, 195011 (2009).

\bibitem{Kubiznak:2012wp}
D.~Kubiznak and R.~B.~Mann,
JHEP \textbf{07}, 033 (2012).

\bibitem{Cai:2013qga}
R.~G.~Cai, L.~M.~Cao, L.~Li and R.~Q.~Yang,
JHEP \textbf{09}, 005 (2013).

\bibitem{Banerjee:2012zm}
R.~Banerjee and D.~Roychowdhury,
Phys. Rev. D \textbf{85}, 104043 (2012).

\bibitem{Li:2018rpk}
S.~L.~Li, H.~D.~Lyu, H.~K.~Deng and H.~Wei,
Eur. Phys. J. C \textbf{79}, no.3, 201 (2019).

\bibitem{Mo:2013ela}
J.~X.~Mo and W.~B.~Liu,
Phys. Lett. B \textbf{727}, 336-339 (2013).

\bibitem{Xu:2015rfa}
J.~Xu, L.~M.~Cao and Y.~P.~Hu,
Phys. Rev. D \textbf{91}, no.12, 124033 (2015).

\bibitem{Momeni:2016qfv}
D.~Momeni, M.~Faizal, K.~Myrzakulov and R.~Myrzakulov,
Phys. Lett. B \textbf{765}, 154-158 (2017).

\bibitem{Karch:2015rpa}
A.~Karch and B.~Robinson,
JHEP \textbf{12}, 073 (2015).

\bibitem{Mancilla:2024spp}
R.~Mancilla,
[arXiv:2410.06605 [hep-th]].

\bibitem{Fernando:2016sps}
S.~Fernando,
Phys. Rev. D \textbf{94}, no.12, 124049 (2016).

\bibitem{Wei:2020rbh}
S.~W.~Wei,
Phys. Rev. D \textbf{102}, no.6, 064039 (2020).

\bibitem{Wei:2022dzw}
S.~W.~Wei, Y.~X.~Liu and R.~B.~Mann,
Phys. Rev. Lett. \textbf{129}, no.19, 191101 (2022).

\bibitem{Wei:2021vdx}
S.~W.~Wei and Y.~X.~Liu,
Phys. Rev. D \textbf{105}, no.10, 104003 (2022).

\bibitem{Yerra:2022alz}
P.~K.~Yerra and C.~Bhamidipati,
Phys. Rev. D \textbf{105}, no.10, 104053 (2022).

\bibitem{Yerra:2022eov}
P.~K.~Yerra and C.~Bhamidipati,
Phys. Lett. B \textbf{835}, 137591 (2022).

\bibitem{Gogoi:2023xzy}
N.~J.~Gogoi and P.~Phukon,
Phys. Rev. D \textbf{108}, no.6, 066016 (2023).

\bibitem{Gogoi:2023qku}
N.~J.~Gogoi and P.~Phukon,
Phys. Rev. D \textbf{107}, no.10, 106009 (2023).

\bibitem{Yerra:2022coh}
P.~K.~Yerra, C.~Bhamidipati and S.~Mukherji,
Phys. Rev. D \textbf{106}, no.6, 064059 (2022).

\bibitem{Wei:2022mzv}
S.~W.~Wei and Y.~X.~Liu,
Phys. Rev. D \textbf{107}, no.6, 064006 (2023).


\bibitem{deMToledo:2018tjq}
J.~de M.Toledo and V.~B.~Bezerra,
Eur. Phys. J. C \textbf{78}, no.7, 534 (2018).

\bibitem{Toledo:2019amt}
J.~M.~Toledo and V.~B.~Bezerra,
Eur. Phys. J. C \textbf{79}, no.2, 110 (2019).

\bibitem{Bekenstein:1973ur}
J.~D.~Bekenstein,
Phys. Rev. D \textbf{7}, 2333-2346 (1973).

\bibitem{Hawking:1974rv}
S.~W.~Hawking,
Nature \textbf{248}, 30-31 (1974).

\bibitem{Hawking:1976de}
S.~W.~Hawking,
Phys. Rev. D \textbf{13}, 191-197 (1976).

\bibitem{Rani:2022xza}
S.~Rani, A.~Jawad, H.~Moradpour and A.~Tanveer,
Eur. Phys. J. C \textbf{82}, no.8, 713 (2022).

\bibitem{Jawad:2022lww}
A.~Jawad and S.~R.~Fatima,
Nucl. Phys. B \textbf{976}, 115697 (2022).

\bibitem{Capozziello:2025axh}
S.~Capozziello and M.~Shokri,
Eur. Phys. J. C 85(\textbf{2}), 200 (2025).

\bibitem{Jawad:2020ihz}
A.~Jawad,
Class. Quant. Grav. \textbf{37}, no.18, 185020 (2020).

\bibitem{Barrow:2020tzx}
J.~D.~Barrow,
Phys. Lett. B \textbf{808}, 135643 (2020).

\bibitem{Barriola:1989hx}
M.~Barriola and A.~Vilenkin,
Phys. Rev. Lett. \textbf{63}, 341 (1989).

\bibitem{SupernovaCosmologyProject:1998vns}
S.~Perlmutter \textit{et al.} [Supernova Cosmology Project],
Astrophys. J. \textbf{517}, 565-586 (1999)

\bibitem{Planck:2018vyg}
N.~Aghanim \textit{et al.} [Planck],
Astron. Astrophys. \textbf{641}, A6 (2020).

\bibitem{Planck:2018jri}
Y.~Akrami \textit{et al.} [Planck],
Astron. Astrophys. \textbf{641}, A10 (2020).

\bibitem{Cardenas:2002np}
R.~Cardenas, T.~Gonzalez, Y.~Leiva, O.~Martin and I.~Quiros,
Phys. Rev. D \textbf{67}, 083501 (2003).

\bibitem{Guo:2004fq}
Z.~K.~Guo, Y.~S.~Piao, X.~M.~Zhang and Y.~Z.~Zhang,
Phys. Lett. B \textbf{608}, 177-182 (2005).

\bibitem{Copeland:2006wr}
E.~J.~Copeland, M.~Sami and S.~Tsujikawa,
Int. J. Mod. Phys. D \textbf{15}, 1753-1936 (2006).

\bibitem{Xia:2006rr}
J.~Q.~Xia, G.~B.~Zhao, H.~Li, B.~Feng and X.~Zhang,
Phys. Rev. D \textbf{74}, 083521 (2006).

\bibitem{Padmanabhan:2007xy}
T.~Padmanabhan,
Gen. Rel. Grav. \textbf{40}, 529-564 (2008).

\bibitem{Durrer:2007re}
R.~Durrer and R.~Maartens,
Gen. Rel. Grav. \textbf{40}, 301-328 (2008).

\bibitem{Bamba:2012cp}
K.~Bamba, S.~Capozziello, S.~Nojiri and S.~D.~Odintsov,
Astrophys. Space Sci. \textbf{342}, 155-228 (2012).

\bibitem{Joyce:2014kja}
A.~Joyce, B.~Jain, J.~Khoury and M.~Trodden,
Phys. Rept. \textbf{568}, 1-98 (2015).

\bibitem{Frusciante:2019xia}
N.~Frusciante and L.~Perenon,
Phys. Rept. \textbf{857}, 1-63 (2020).





\bibitem{Banks:1998vs}
T.~Banks, W.~Fischler and L.~Motl,
JHEP \textbf{01}, 019 (1999).

\bibitem{Hellerman:2001yi}
S.~Hellerman, N.~Kaloper and L.~Susskind,
JHEP \textbf{06}, 003 (2001).

\bibitem{Kiselev:2002dx}
V.~V.~Kiselev,
Class. Quant. Grav. \textbf{20}, 1187-1198 (2003).

\bibitem{Liang:2020hjz}
J.~Liang, X.~Guo, D.~Chen and B.~Mu,
Nucl. Phys. B \textbf{965}, 115335 (2021).

\bibitem{Chabab:2020ejk}
M.~Chabab and S.~Iraoui,
Gen. Rel. Grav. \textbf{52}, no.8, 75 (2020).

\bibitem{Ma:2019pya}
Y.~Ma, Y.~Zhang, R.~Zhao, S.~Cao, T.~Liu, S.~Geng, Y.~Liu and
Y.~Huang,
Int. J. Mod. Phys. D 29(\textbf{15}), 2050108 (2020). 

\bibitem{Davies:1989ey}
P.~C.~W.~Davies,
Class. Quant. Grav. \textbf{6}, 1909 (1989).

\bibitem{Lousto:1994jd}
C.~O.~Lousto,
Phys. Rev. D \textbf{51}, 1733-1740 (1995).

\bibitem{Muniain:1995ih}
J.~P.~Muniain and D.~D.~Piriz,
Phys. Rev. D \textbf{53}, 816-823 (1996).

\bibitem{Bhattacharya:2024bjp}
K.~Bhattacharya, K.~Bamba and D.~Singleton,
Phys. Lett. B \textbf{854}, 138722 (2024).


\bibitem{Bhattacharya:2021lgk}
K.~Bhattacharya,
Nucl. Phys. B \textbf{989}, 116130 (2023).

\bibitem{Ladghami:2024sen}
Y.~Ladghami, B.~Asfour, A.~Bouali, A.~Errahmani and T.~Ouali,
Phys. Dark Univ. \textbf{44}, 101470 (2024).

\bibitem{Duan:1979ucg}
Y.~S.~Duan and M.~L.~Ge,
Sci. Sin. \textbf{9}, no.11, 1072 (1979).

\bibitem{Duan:1984ws}
Y.~Duan,
``THE STRUCTURE OF THE TOPOLOGICAL CURRENT,''
SLAC-PUB-3301.

\end{thebibliography}
\end{document}